\documentclass[a4paper]{jpconf}
\usepackage{graphicx}
\usepackage{amsmath}	
\usepackage{amssymb}	
\usepackage[sort, numbers]{natbib}

\def\beq{\begin{equation}}
\def\eeq{\end{equation}}
\def\bea{\begin{eqnarray}}
\def\eea{\end{eqnarray}}

\begin{document}
\title{Constraining the charge of \\the Galactic centre black hole}

\author{Michal Zaja\v{c}ek$^{1,2,3}$, Arman Tursunov$^{2,4}$, Andreas Eckart$^{2,1}$, Silke Britzen$^{1}$, Eva Hackmann$^{5}$, Vladim\'ir Karas$^{6}$, Zden\v{e}k Stuchl\'ik$^{4}$, Bozena Czerny$^{3}$, and J. Anton Zensus$^{1}$}

\address{$^{1}$Max-Planck-Institut f\"ur Radioastronomie (MPIfR), Auf dem H\"ugel 69, D-53121 Bonn, Germany}
\address{$^{2}$I. Physikalisches Institut der Universit\"at zu K\"oln, Z\"ulpicher Strasse 77, D-50937 K\"oln, Germany}
\address{$^{3}$Center for Theoretical Physics, Polish Academy of Sciences, Al. Lotnikow 32/46, 02-668 Warsaw, Poland}
\address{$^{4}$Institute of Physics and Research Centre of Theoretical Physics and Astrophysics, Faculty of Philosophy and Science,\\Silesian University in Opava, Bezru\v{c}ovo n\'{a}m.13, CZ-74601 Opava, Czech Republic}
\address{$^{5}$ZARM, University of Bremen, Am Fallturm, 28359 Bremen, Germany}
\address{$^{6}$Astronomical Institute, Academy of Sciences, Bo\v{c}n\'{\i}~II 1401, CZ-14131~Prague, Czech Republic}

\ead{zajacek@ph1.uni-koeln.de}

\begin{abstract}
In this contribution, we summarize our results concerning the observational constraints on the electric charge associated with the Galactic centre black hole -- Sgr~A*. According to the no-hair theorem, every astrophysical black hole, including supermassive black holes, is characterized by at most three classical, externally observable parameters -- mass, spin, and the electric charge. While the mass and the spin have routinely been measured by several methods, the electric charge has usually been neglected, based on the arguments of efficient discharge in astrophysical plasmas. From a theoretical point of view, the black hole can attain charge due to the mass imbalance between protons and electrons in fully ionized plasmas, which yields about $\sim 10^8\,{\rm C}$ for Sgr~A*. The second, induction mechanism concerns rotating Kerr black holes embedded in an external magnetic field, which leads to electric field generation due to the twisting of magnetic field lines. This electric field can be associated with the induced Wald charge, for which we calculate the upper limit of $\sim 10^{15}\,{\rm C}$ for Sgr~A*. Although the maximum theoretical limit of $\sim 10^{15}\,{\rm C}$ is still 12 orders of magnitude smaller than the extremal charge of Sgr~A*, we analyse a few astrophysical consequences of having a black hole with a small charge in the Galactic centre. Two most prominent ones are the effect on the X-ray bremsstrahlung profile and the effect on the position of the innermost stable circular orbit.    
\end{abstract}

\section{Sgr~A* as a supermassive black hole}
The compact radio source Sgr~A* was detected by B. Balick and R. Brown \cite{1974ApJ...194..265B} in 1974 using the 35-km baseline interferometer Green Bank-Huntersville. Its brightness temperature was in excess of $10^7\,{\rm K}$ and the source angular size was $\lesssim 0.1''$. They measured the flux density at two frequencies, $2.7\,{\rm GHz}$ and $8.1\,{\rm GHz}$ - $S_{2.7}\approx 0.6\,{\rm Jy}$ and $S_{8.1}\approx 0.8\,{\rm Jy}$. The compact radio source was located at the very centre ($\sim 2-3\,{\rm pc}$ linear scale) of the brightest IR/radio complex of Sgr A \cite{1971Natur.233..112D,1973ApJ...184..415R,1974ApJ...192..325B}. The large brightness temperature, compactness, and the association of the radio source with the centre of the Galaxy were consistent with the black-hole hypothesis of Donald Lynden-Bell and Martin Rees \cite{1969Natur.223..690L,1971MNRAS.152..461L}. These properties were confirmed by the Very Long Baseline Interferometry observations in 1975 by Fred Lo et al. \cite{1975ApJ...202L..63L}, who resolved the compact radio source Sgr~A* up to $0.02''$ at the wavelength of $3.7\,{\rm cm}$ ($8.1\,{\rm GHz}$) using the baseline of $242\, {\rm km}$. They compared their observations with the previous ones, which implied the source variability. Brown et al. \cite{1981ApJ...250..155B} used the Very Large Array (VLA) at 5 GHz to make a radio map of the central region with the angular resolution $(2''\times 8'')=(\alpha \times \beta)$ comparable with the infrared 10 ${\rm \mu m}$ maps of the same region \cite{1978ApJ...219..121B,1978ApJ...220..831B}. The radio-infrared comparison made it possible to place unresolved non-thermal source Sgr~A* at the dynamical centre of the motion of thermal streamers in Sgr~A~West, whose kinematics was inferred based on $12.8\,{\rm \mu m}$ fine-structure emission of NeII \cite{1976ApJ...205L...5W,1977ApJ...218L.103W,1979ApJ...227L..17L,1980ApJ...241..132L}. 

Robert Brown \cite{1982ApJ...262..110B} introduced the designation Sgr~A* for the first time to make a distinction between the compact radio source located in an extended, complex radio emission of Sgr~A. Further confirmation of the central mass of $\sim 3\times 10^6\,M_{\odot}$ came with the analysis of neutral 63 ${\rm \mu m}$ fine-structure emission of oxygen [OI] in combination with NeII emission \cite{1984ApJ...276..551G}. Enough evidence had been accumulated that the central parsec hosts the largest concentration of stars in the Galaxy that co-exist with the ionized gas and warm dust as well as a massive compact object, presumably a black hole, at the very centre \cite{1982ApJ...262..120L,1983Natur.301..661T}. The intrinsic variability of Sgr~A* was confirmed with dual-frequency observations ($2.7$ and $8.1$ GHz) during 25 epochs over the period of 3 years \cite{1982ApJ...253..108B}. The flux density varied by $20\%$-$40\%$ on the timescales from days to years. 

Consequently, the variability of Sgr~A* was linked to a down-scaled quasar activity and the source become one of the prime candidates for a supermassive black hole. One of the ways to test the black-hole hypothesis is to constrain a mass concentration within a certain volume or the mean density. For a black hole of mass $M_{\bullet}$, the mean density can be derived within the volume given by the Schwarzschild radius $r_{\rm s}=2GM_{\bullet}/c^2=2r_{\rm g}$,

\begin{equation}
    \rho_{\bullet}=\frac{3c^6}{32\pi G^3M_{\bullet}^2}=1.71 \times 10^{25}\left(\frac{M_{\bullet}}{4\times 10^6\,M_{\odot}} \right)^{-2}\,{\rm M_{\odot}\,pc^{-3}}\,.
    \label{eq_density_bh}
\end{equation}

Thanks to the long-term monitoring of S2 star, which passed the pericentre in May 2018 at the distance of $120\,{\rm AU}\approx 1400\,r_{\rm s}$ \citep{2018A&A...615L..15G_new}, it is possible to constrain the mean density for the fitted mass of $M_{\bullet}=(4.100 \pm 0.034) \times 10^6\,M_{\odot}$ and the pericentre distance of $r_{\rm p} \simeq 0.577\,{\rm mpc}$: $\rho_{\rm S2}\simeq 5.1 \times 10^{15}\,{\rm M_{\odot}\,yr^{-3}}$.

A significant improvement came with the first detection of orbital motions near the innermost stable circular orbit (ISCO) of Sgr~A* \cite{2018A&A...618L..10G}. The motion occurred during the bright phases of Sgr~A* activity -- so called flares -- and the NIR observations are consistent so far with the hot-spot interpretation of NIR/X-ray flares. The hot spot motion in the strong-gravity regime was modelled previously and fitted to NIR and X-ray light curves \cite{2005MNRAS.363..353B,2006ApJ...636L.109B,2006A&A...460...15M,2017MNRAS.472.4422K}. The detection of a hot spot by the GRAVITY instrument \cite{2008poii.conf..431E,2019hsax.conf..609A} at the Very Large Telescope Interferometer (VLTI) facility of the European Southern Observatory (ESO) on Paranal shows both positional and polarization changes on the timescale of $45(\pm 15)\,{\rm min}$ and is consistent with the nearly face-on clockwise motion of a nonthermal hot spot close to the ISCO of $\sim 4\times 10^6\,M_{\odot}$ black hole ($6-10\,GM_{\bullet}/c^2$). This puts the lower limit on the mean mass density of enclosed matter, $\rho_{\rm HS}>1.3 \times 10^{23}\,M_{\odot}{\rm pc^{-3}}$, which is only two orders of magnitude smaller than the density expected for a black hole, see Eq.~\eqref{eq_density_bh}. Essentially the same order of magnitude for the mass density is given by the detection of the compact intrinsic structure of Sgr~A* on the length-scale of 3 Schwarzschild radii, which was obtained by the VLBI technique at $1.3$ mm \cite{2018ApJ...859...60L}. The detected spatial structure associated with the mass of $4\times 10^6\,M_{\odot}$ then yields $\rho_{\rm VLBI}\approx 6 \times 10^{23}\,M_{\odot}{\rm pc^{-3}}$. 

In addition to observationally confirmed compactness of Sgr~A* on the scale of 3 Schwarzschild radii and its intrinsic multiwavelength variability, two general relativistic predictions were confirmed within uncertainties: the pericentre shift for S2 star \cite{2017ApJ...845...22P} and the combined gravitational and transverse Doppler shift \cite{2018A&A...615L..15G_new}. Hence, there is currently only a little space for non-black-hole hypotheses concerning the nature of Sgr~A*, which have a similar degree of compactness as black holes (gravastar, boson star, fermion ball, fuzzball, holostars, dark stars to name a few alternatives, see \cite{2017FoPh...47..553E} for an extensive discussion) with their surface located just above the event horizon with the radial difference of small $f$. The main differences between classical black holes and the majority of alternatives is that they miss both the event horizon and the singularity, which is induced by their quantum nature. The presence of the event horizon instead of the solid surface was supposed to be supported by the non-detection of the X-ray thermal emission or thermal flares. In particular, the viscously dissipated energy is advected through the event horizon, which can explain the very low radiative efficiency \cite{1998ApJ...492..554N}. However, the fact that black-hole candidates appear dimmer and/or lack the thermal component cannot be by itself used as a proof of the black hole hypothesis as the extremely compact quantum states would exhibit essentially the same radiative properties for a distant observer \cite{2002A&A...396L..31A,2017FoPh...47..553E}. Although the gravitational-wave detection provides a significant improvement in determining the nature of merging compact objects, it is still quite a challenge to make definite conclusions based on the detected ringdown waveforms as these manifest the presence of light rings rather than of horizons \cite{2016PhRvL.116q1101C}. 
On the other hand, the quasi-normal modes for black holes and various alternatives in general are different, which allows one to exclude some of the alternatives by comparing  signals of gravitational wave events with corresponding ringdown waveforms of compact objects \cite{2007CQGra..24.4191C,2016PhRvD..94h4016C}.
It is relevant to note that a general problem of observational tests of black holes is even more complex. First, there are alternatives to classical black holes with event horizons but without a singularity,
like the regular black holes which can be constructed by coupling
general relativity to a non-linear electrodynamics \cite{Bardeen1968, Ayon-Beato1998}. Second, also naked singularity solutions could be potentially relevant, even in the scope of general relativity and observational astrophysics \cite{PhysRevD.76.064024}. 

Applying Occam's razor, in the following we consider Sgr~A* to be described by a Kerr-Newman black hole, which is the most general solution with general relativity, to which the no-hair theorem applies, i.e. it is fully described by three classical, externally observable paramters: mass, spin, and an electric charge.

\subsection{Mass determination of Sgr~A*}
The first estimate of the mass of the compact radio source Sgr~A* was based on the spatially and spectrally resolved NeII fine-structure emission ($12.8\,{\rm \mu m}$) within the region of $40''$ coincident with the thermal radio-continuum region of Sgr~A~West \cite{1976ApJ...205L...5W,1977ApJ...218L.103W}. The Doppler-shifted velocities of NeII line ranged from $+250\,{\rm km\,s^{-1}}$ to $-350\,{\rm km\,s^{-1}}$ with respect to the LSR. Based on velocities and velocity dispersions, Wollmann et al. \cite{1976ApJ...205L...5W,1977ApJ...218L.103W} inferred that the enclosed mass with the inner parsec is $\sim 4\times 10^6\,M_{\odot}$. This is essentially consistent with the current value based on the Newtonian, simultaneous orbital fits to three stars (S2, S38, and S55/S0-102) in the S cluster by Parsa et al. \cite{2017ApJ...845...22P}, who obtained the mass $M_{\bullet}=(4.15 \pm 0.13 \pm 0.57) \times 10^6\,M_{\odot}$ as well as the distance $R_{\bullet}=8.19 \pm 0.11 \pm 0.34\,{\rm kpc}$. The highest angular resolution observations using the Very Large Telescope Interferometer facility at the European Southern Observatory on the Paranal mountain -- GRAVITY \cite{2008poii.conf..431E,2019hsax.conf..609A} -- in the near-infrared $K_{\rm s}$ band ($2.2\,\mu{\rm m}$) were employed to obtain the mass as well as the distance to the Galactic centre using the best-fit orbit of the star S2 with and without Schwarzschild precession with both sets of values to be comparable \cite{2018A&A...615L..15G_new}. Considering Schwarzschild precession, the best-fit values are $M_{\bullet}=(4.100 \pm 0.034) \times 10^6\,M_{\odot}$ and $R_{\bullet}=8.122 \pm 0.031\,{\rm kpc}$.

\subsection{Spin determination of Sgr~A*}
\label{spin_determination}

The determination of the spin value for Sgr~A* is a model-dependent quantity. In particular, it relies on the comparison of flare data (light curves and orbital evolution) with general relativistic model predictions. Sgr~A* exhibits flares across the whole electromagnetic spectrum whose periodicity of $\sim 17\,{\rm min}$ has been employed to infer the spin of Sgr~A* \cite{2003Natur.425..934G,2006A&A...460...15M,2012A&A...537A..52E}, which was done by identifying the flare period with the period corresponding to the innermost stable circular orbit (ISCO). Near-infrared and X-ray flares seem to originate just outside the event horizon and the inferred periodicity is consistent with the spin of $a_{\bullet} \sim 0.5$ \cite{2003Natur.425..934G} (half of extremal value). Recently, Witzel et al. \cite{2018ApJ...863...15W} constrain the spin to even higher values, $a_{\bullet}>0.9$, based on the complex statistical analysis of NIR flares. In particular, they do not find any structure in the power spectral density below $<8.5$ minutes, which is the timescale associated with the ISCO of a high-spinning black hole of $\sim 4 \times 10^6\,M_{\odot}$. In general, the observed light curves exhibit modulations, with the indication of shortening of the period, which implies the plunging of the emitting matter inwards. The spin determination depends primarily on the position of the innermost or marginally stable circular orbit (ISCO). However, when magnetohydrodynamic interaction is included, the \textit{stress edge radius} becomes important, which is the radius at which the orbiting matter becomes dynamically decoupled from the rest of the accretion flow further out \cite{2003bhcg.book.....M}. The stress edge radius in MHD simulations was found to be smaller than the ISCO, about $r_{\rm stress} \sim 2.2-2.3\,r_{\rm s}$ in comparison with $r_{\rm ISCO}=3\,r_{\rm s}$ for a non-rotating black hole. Last but not the least, in case the black hole is charged, as we investigate below, then the charge could mimick the spin for charged particles in terms of shifting the ISCO, which can make the determination of the spin of Sgr~A* even more complex (see \cite{2018MNRAS.480.4408Z} for further details).   

\subsection{Charge determination of Sgr~A*}

In most studies of Sgr~A*, it is usually assumed that the black hole is uncharged ($Q_{\bullet}\equiv0$). This is backed up by the argumentation that any charge of Sgr~A* would be quickly discharged by the inflow of particles of opposite charge from the plasma in the vicinity. On the other hand, the rotating black hole immersed in the external magnetic field with a non-negligible poloidal component leads to the induced \textit{Wald charge} \cite{1974PhRvD..10.1680W}. In Subsection~\ref{spin_determination}, we listed the arguments that the supermassive black hole associated with Sgr~A* has a non-zero spin. Moreover, the presence of the dominantly poloidal magnetic field close to Sgr~A* is supported by the recent detection of the orbital motion -- orbiting \textit{hot spot} \cite{2018A&A...618L..10G}. Hence, there seems to be a discrepancy between the usual assumption of zero charge and the theoretical expectations, which in general imply non-zero values of electric charge.

In the following discussion, it is more precise to use the term \textit{electromagnetized Kerr black hole}, as the astrophysical electromagnetic fields are too weak to significantly
influence the Kerr spacetime geometry. They are, however, strong enough to
have an impact on charged matter. The electromagnetic fields can be
external, i.e. not directly associated with the black hole, or
internally arising from the electric charge of the black hole. There is
a fundamental difference between the electromagnetized Kerr black hole
and the KN black hole -- in relation to the test particle motion that is completely regular in the KN black-hole case, but it is of a chaotic origin (deterministic chaos) in the case of magnetized black holes \cite{2014ApJ...787..117K}, especially when the axial symmetry is significantly perturbed e.g. by an oblique external magnetic field. The motion of charged particles around magnetized black holes has been analyzed in several studies \cite{kovar2010,kopacek2010,2015CQGra..32p5009K,2016PhRvD..93h4012T,Stuchlik2016,Kolos2017,2018ApJ...861....2T}. The space-time structure of charged black holes was investigated in detail as well \cite{1991JPhy1...1.1005K,1991JMP....32..714K,2011PhRvD..84h4002K,2011PhRvD..83j4052P}.

The effect of the small electric charge, i.e. negligible in terms of the space-time metric, on the accretion of plasma was investigated in \cite{2017PhRvD..96f3015S} for a rotating and charged black hole. They showed that the small charge can significantly effect the plasma motion in case the electromagnetic field of the plasma is small. On the theoretical level, charged equilibrium
tori as a toy model for charged accretion disks around an
electromagnetized Kerr black hole were analysed in \cite{2018PhRvD..97j4019T} and it was found that the spin can affect the equilibrium conditions in a significant way in comparison with the set-up of charged accretion tori around non-rotating charged black holes \cite{2011PhRvD..84h4002K}; see also the original works about the properties of \textit{dielectric tori} around compact objects: \cite{2011PhRvD..84h4002K,slany2013,cremaschini2013,kovar2014}.

In the further discussion and analysis, we look at both the constraints on the charge of Sgr~A* that stem from fundamental principles and on the observational implications, namely how to observationally distinguish between the charged and the non-charged black hole associated with Sgr~A*. Previously, based on the VLBI observations of Sgr~A* and the motion of S2 star, the constraints on the charge of Sgr~A* were rather weak and in general associated with the extremal charge \cite{2012GReGr..44.1753I,2014PhRvD..90f2007Z}. In \cite{2018MNRAS.480.4408Z}, we tried to put tighter constraints on the electric charge of Sgr~A*, which is also summarized in the following Sections. 

\section{Theoretical constraints on the charge of Sgr A*}

\begin{figure}[h!]
    \centering
    \includegraphics[width=\textwidth]{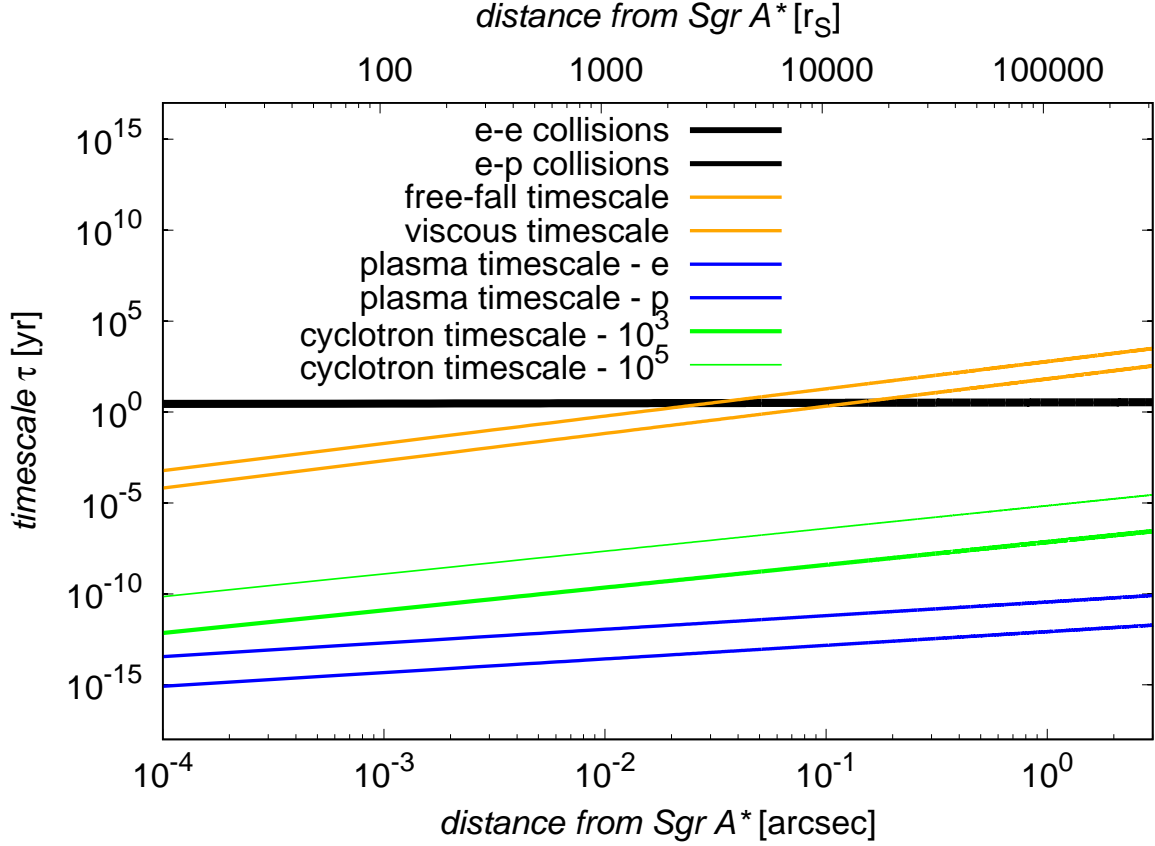}
    \caption{Comparison of typical timescales (expressed in years) of the accretion flow (free-fall and viscous timescales) with the timescales that characterize the plasma close to Sgr~A* (electron-electron and electron-proton collisions, timescales derived from the plasma and the cyclotron frequency). The bottom $x$-axis expresses the distace from Sgr~A* in arcseconds, the upper axis in Schwarzschild radii.}
    \label{fig_timescales}
\end{figure}

The charge of Sgr~A*, if present, is directly linked to magnetohydrodynamic environment of the Galactic centre. The observations of the vicinity of Sgr~A* in X-ray inside the Bondi radius \cite{2013Sci...341..981W},

\begin{equation}
  R_{\rm B}\approx 0.125 \left(\frac{M_{\bullet}}{4\times 10^6\,M_{\odot}}\right) \left(\frac{T_{\rm e}}{10^7\,{\rm K}} \right)^{-1}\left(\frac{\mu_{\rm HII}}{0.5}\right)\,{\rm pc}\,,
  \label{eq_Bondi_radius}
\end{equation}

reveal hot plasma of $\sim 10^7\,{\rm K}$ that emits thermal bremsstrahlung \cite{2003ApJ...591..891B,2013Sci...341..981W,2015A&A...581A..64R}. In addition, this plasma is very diluted and is therefore very weakly coupled. The coupling parameter $R_{\rm c}$ may be expressed as the ratio of the potential to kinetic energy,

\begin{equation}
 R_{\rm c}=\frac{E_{\rm p}}{E_{\rm k}}\sim \frac{e^2(L_{\rm i} 4\pi \epsilon_0)^{-1}}{k_{\rm B}T_{\rm e}}=  \frac{e^2n_{\rm p}^{1/3}(4\pi \epsilon_0)^{-1}}{k_{\rm B}T_{\rm e}}\simeq 10^{-10}\,,
\end{equation}
where $L_{\rm i}$ is the mean interparticle distance, which can be estimated from the mean particle density, $L_{\rm i}\sim n_{\rm p}^{-1/3}$, and $T_{\rm e}$ is the electron temperature. For the inferred particle density at the Bondi radius $R_{\rm B}$, $n_{\rm p}\sim n_{\rm e}\approx 10\,{\rm cm^{-3}}$ \cite{2003ApJ...591..891B,2013Sci...341..981W}, and the electron temperature $k_{\rm B}T_{\rm e}\sim 1\,{\rm keV}$, we get the coupling parameter $R_{\rm c}\sim 3 \times 10^{-10}$, which implies the very small coupling of the Galactic centre plasma, i.e. it may be treated as collisionless. This is also apparent in Fig.~\ref{fig_timescales}, where we compare characteristic electron-electron and electron-proton collisional timescales with dynamical timescales of the accretion flow. Inside the inner 1000 Schwarzschild radii, typical collisional timescales are longer than the dynamical timescales. On the other hand, the characterstic ordered charge oscillations of the plasma (plasma frequency) and cyclotron motions take place on much shorter timescales than the viscous or free-fall timescale. For details, see the analysis presented in \cite{2018MNRAS.480.4408Z}. 

\subsection{Classical considerations of charging}

In the first approximation, we will treat Sgr~A* as a massive object of mass $M_{\bullet}$ that is immersed in the fully ionized stationary plasma composed of protons and electrons that are not coupled to each other. In such an atmosphere, lighter electrons will tend to separate from heavier protons. Such a separation will continue until a charge $Q_{\bullet}$ associated with the black hole is induced that stops further separation. The total potential energy of electrons can be expressed as, $W_{\rm e}=e\phi-e\theta$, and the potential energy of protons is, $W_{\rm p}=e\phi+e\theta$, where $\phi=-GM_{\bullet}/r$ is the gravitational potential and $\theta=Q_{\bullet}/(4 \pi \epsilon_0 r)$ is the electrostatic potential. In the equilibrium configuration of plasma, the number densities of electrons and protons are expressed by the Maxwell-Boltzmann statistical distribution, $n_{\rm e}\propto \exp{(-W_{\rm e}/k_{\rm B}T_{\rm e})}$ and $n_{\rm p}\propto \exp{(-W_{\rm p}/k_{\rm B}T_{\rm p})}$, respectively. In the ionized plasma around Sgr~A*, normally we expect a quasineutral plasma with $n_{\rm e}\sim n_{\rm p}$, which implies $W_{\rm e}\sim W_{\rm p}$. This leads to the estimate of the equilibrium induced charge $Q_{\bullet}^{\rm eq}$:

\begin{align}
  Q_{\bullet}^{\rm eq}  &= \frac{2\pi \epsilon_0 G(m_{\rm p} - m_{\rm e})}{e}M_{\bullet}\,\notag\\                 &\approx 3.1 \times 10^8 \left(\frac{M_{\bullet}}{4\times 10^6\,M_{\odot}} \right)\,C\,,
  \label{eq_charge_blackhole}
\end{align}
which gives the charge to mass ratio for Sgr~A*, $Q_{\bullet}^{\rm eq}/M_{\bullet}\approx 76\, C/M_{\odot}$, which is essentially the same as originally derived by A. S. Eddington for the Sun \cite{1926ics..book.....E,2001A&A...372..913N}. John Bally and E. R. Harrison \cite{1978ApJ...220..743B} generalized the analysis for all massive bodies surrounded by plasma including galaxies and concluded that they are positively charged with the charge to mass ratio of $\sim 100\,{\rm  C}/M_{\odot}$. Moreover, this positive charge is not screened by negative electron gas due to the large length-scale of the objects in comparison with the Debye length $\lambda_{\rm D}$. For the Galactic centre, we get an estimate of the Debye length at the ISCO using the density and temperature profiles from \cite{2015A&A...581A..64R},
\begin{equation}
  \lambda_{\rm ISCO}=5\,\left(\frac{T_{\rm e}}{8.7\times 10^{12}\,{\rm K}} \right)^{1/2}\left(\frac{n_{\rm e}}{1.7\times 10^9\,{\rm cm^3}} \right)^{-1/2}\,{\rm m}\,,
  \label{eq_Debye_Bondi}
\end{equation}
and at the Bondi radius, where densities are significantly lower, it will be only one order of magnitude larger, $\lambda_{\rm Bondi}\approx 141\,{\rm m}$. The length-scale of Sgr~A* is given by its Schwarzschild radius, $r_{\rm s}=2GM_{\bullet}/c^2=1.18 \times 10^{10}\,{\rm m} \gg \lambda_{\rm Bondi}> \lambda_{\rm ISCO}$. Hence, the charge associated with the Galactic centre black hole is not entirely screened by a negatively charged hot accretion flow since the length-scale of Sgr~A* and essentially all supermassive black holes is much bigger than the Debye length-scale as derived for the temperature and the density of the medium in its surroundings. According to \cite{1978ApJ...220..743B}, any self-gravitating medium, whose length-scale is larger than the Debye length of the surrounding ISM atmosphere, is expected to be positively charged. This is certainly the case for Sgr~A* plus the surrounding hot flow up to the Bondi radius.

\subsection{General Relativistic estimates of electric charge}

The supermassive black hole at the Galactic centre can be treated in the most general case as a Kerr-Newman (KN) black hole \cite{1963PhRvL..11..237K,1965JMP.....6..918N}. The extremal KN black hole has a single event horizon for the extremal charge of

\begin{equation}
  Q_{\rm max}^{{\rm rot}}=2M_{\bullet}\sqrt{\pi \epsilon_0 G (1-\tilde{a}_{\bullet}^2)}\,,
  \label{eq_max_gen_nodim}
\end{equation}
which for the non-rotating case, $\tilde{a}_{\bullet}=0$,  may simply be evaluated as
\begin{equation}
 Q_{\rm max}^{{\rm norot}}=2\sqrt{\pi \epsilon_0 G}M_{\bullet}=6.86 \times 10^{26}\, \left(\frac{M_{\bullet}}{4\times 10^6\,M_{\odot}} \right)\,C\,.
 \label{eq_max_charge}
\end{equation} 

The following analysis and estimates are based on the fact that Sgr~A* has a non-zero spin and is surrounded by a magnetohydrodynamic medium (plasma$+$magnetic field), hence it is not located in the vacuum \cite{2010RvMP...82.3121G}. This is not in contradiction with classical solutions of Einstein field equations, which assume bodies to be located in vacuum, since the mass of the plasma is negligible with the mass of Sgr~A* and typically, the plasma dynamics is treated separately using hydrodynamic and Maxwell equations in the background of Kerr or Kerr-Newman metric. In case of self-gravitating structures such as discs, one could include relevant perturbative terms into Kerr-Newman metric \cite{2004CQGra..21R...1K,2010MNRAS.404..545S,2012MNRAS.425.2455S,2013MNRAS.436..978S,2015MNRAS.451.1770W}. 

Previously, we provided evidence that the black hole has a non-zero spin parameter, $a_{\bullet}\gtrsim 0.4$ (see also \cite{2017FoPh...47..553E} for a review). In addition, it is also immersed in the external magnetic field with the magnitude of $B_{\rm ext}\sim 10 - 100\,{\rm G}$ as inferred from the flare analysis \cite{2012A&A...537A..52E}. The magnetic field in the Galactic centre is highly ordered on the larger scales of $10-100\,{\rm pc}$ \cite{2015llg..book..391M}. On the length-scale of the ISCO, it appears to have a strong poloidal component with respect to the recently detected orbiting hot spot, which was inferred from the detected polarized NIR emission \cite{2018A&A...618L..10G}. The mm-VLBI observations of Sgr~A* detected linearly polarized emission at $1.3\,{\rm mm}$, which implies ordered magnetic field on event-horizon scales with a characteristic intra-hour variability timescale \cite{2015Sci...350.1242J}. Under the assumption that the hot spot orbits the black hole close to its equatorial plane, Sgr~A* is immersed in the strong, ordered poloidal magnetic field, which has direct implications for its electric charge since this is basically a model set-up as analyzed originally by Wald \cite{1974PhRvD..10.1680W}.

The electric field is generated by twisting magnetic field lines of the circumnuclear magnetic field with a strong poloidal component due to the black hole rotation. The magnetic field is expected to possess the properties of the background space-time metric: axial symmetry and stationarity. Then the four-vector potential $A^{\mu}$ may be expressed as the linear combination of the corresponding Killing vectors related to space-time symmetries, $A^{\mu}=k_1\xi_{(t)}^{\mu}+k_2\xi_{(\phi)}^{\mu}$, where $k_1$ and $k_2$ are the constants to be determined by solving Maxwell equations, which yields \cite{1974PhRvD..10.1680W},

\beq 
A_t = \frac{B}{2} \left(g_{t\phi} + 2 a g_{tt}\right), \quad A_{\phi} =  \frac{B}{2} \left(g_{\phi\phi} + 2 a g_{t\phi}\right).
\label{VecPotTP}
\eeq

The black hole rotation leads essentially to the Faraday induction, where the time component of the four-potential $A_{t}$ represents the induced electric field. A potential difference $\Delta \phi$ between the black-hole horizon and the infinity may be evaluated as,
\beq
\Delta \phi = \phi_{\rm H} - \phi_{\infty} = \frac{Q_{\bullet} - 2 a_{\bullet} M_{\bullet} B_{\rm ext}}{2 M_{\bullet}}\,,
\label{eq_potential_ddiference}
\eeq
which leads to the selective accretion of charges from the surrounding plasma until the potential difference is zero. Hence, the maximum net charge can be obtained directly from Eq.~\ref{eq_potential_ddiference}, $Q_{\bullet}^{\rm max}=2a_{\bullet}M_{\bullet}B_{\rm ext}$. Specifically, for Sgr~A*  we constrain the upper limit for the induced charge based on the maximum rotation, $a_{\bullet}\leq M_{\bullet}$,
 \beq 
Q_{\bullet \rm ind}^{\rm max} = 2.32 \times 10^{15} \left( \frac{M_{\bullet}}{4 \times 10^6 M_{\odot}} \right)^2  \left( \frac{B_{\rm ext}}{10 \rm G} \right)  ~\rm C.
\label{eq_max_induced_charge}
\eeq

In Table~\ref{tab_summary_constraints}, we summarize both the classical and relativistic estimates of the Galactic centre black hole. There are two further electrostatic barriers where the accretion of protons and electrons is stopped in the classical limit, 

\beq 
Q_{\rm max}^{+}=6.16\times 10^8\,\left(\frac{M_{\bullet}}{4\times 10^6\,M_{\odot}} \right)\,\rm C, \quad  Q_{\rm max}^{-}=3.36\times 10^5\,\left(\frac{M_{\bullet}}{4\times 10^6\,M_{\odot}} \right)\,\rm C.
\label{eq_electrostatic_barriers}
\eeq

The electrostatic barriers~\eqref{eq_electrostatic_barriers} are, however, not absolute in the general relativistic calculations, as they depend on the radial coordinate $r$ in the following way \cite{2018MNRAS.480.4408Z},

\beq \label{Q-equil-SI}
Q_{\rm max}^{\rm rel} = 4 \pi \epsilon_0 G M_{\bullet} \frac{m_{\rm par}}{q_{\rm par}} \left(1-\frac{r_{\rm s}}{r}\right)^{-1/2}=Q^{+/-}_{\rm max}\left(1-\frac{r_{\rm s}}{r}\right)^{-1/2},
\eeq
where the factor $\left(1-\frac{r_{\rm s}}{r}\right)^{-1/2}$ is the general relativistic correction to the Newtonian limits $Q^{+/-}_{\rm max}$ listed in Table~\ref{tab_summary_constraints}. Relation~\eqref{Q-equil-SI} implies that the black hole charge needs to approach the infinite value to keep the charged particle at the equilibrium position close to the event horizon. The previous analysis shows that the electric charge in the interval $Q_{\bullet}^{\rm Sgr~A*}=(10^8, 10^{15})\,{\rm C}$ is plausible for the Galactic centre black hole. The charge values are at least twelve orders of magnitude smaller than the extremal value of $Q_{\rm max}<10^{27}\,{\rm C}$, which implies that the background metric is not affected. In other words, the motion of neutral bodies is not influenced by the small, likely positive electric charge of Sgr~A*. However, the dynamics of charged particles in the plasma can be profoundly impacted, which is discussed in more detail in the following Section.    

\begin{table} 
  \centering
  \caption{Summary of the constraints for electric charge as evaluated for Sgr~A* supermassive black hole.}
  \resizebox{\textwidth}{!}{  
  \begin{tabular}{c|c|c}
  \hline
  \hline
     \textbf{Process} & \textbf{Limit} & \textbf{Notes}\\
  \hline
     \textbf{Mass difference between $p$ and $e$} & $Q_{\rm eq}=3.1 \times 10^8\,\left(\frac{M_{\bullet}}{4\times 10^6\,M_{\odot}} \right){\rm C}$ & stable charge \\
     \textbf{Accretion of protons}   & $Q_{\rm max}^{+}=6.16\times 10^8\,\left(\frac{M_{\bullet}}{4\times 10^6\,M_{\odot}} \right)\,\rm C$    & electrostatic barrier  \\
     \textbf{Accretion of electrons} & $ Q_{\rm max}^{-}=3.36\times 10^5\,\left(\frac{M_{\bullet}}{4\times 10^6\,M_{\odot}} \right)\,\rm C$   & electrostatic barrier   \\
     \textbf{Magnetic field \& SMBH rotation}  & $Q_{\bullet \rm ind}^{\rm max} \lesssim 10^{15} \left( \frac{M_{\bullet}}{4 \times 10^6 M_{\odot}} \right)^2  \left( \frac{B_{\rm ext}}{10 \rm G} \right)~  \rm C$     & stable charge\\
       \hline
     \textbf{Extremal SMBH} & $Q_{\rm max}=6.86 \times 10^{26}\, \left(\frac{M_{\bullet}}{4\times 10^6\,M_{\odot}} \right)\sqrt{1-\tilde{a}_{\bullet}^2}\,\rm C$  & extremal charge \\
     \hline
  \end{tabular}
  }
  \label{tab_summary_constraints}
\end{table}

\section{Observational consequences}

Since the value of the electric charge is likely at least twelve orders of magnitude below the extremal value, observational tests to distinguish charged black holes from non-charged ones must primarily involve charged elementary particles. The effect on neutral bodies and photons is apparent only for values close to the extremal charge.

\subsection{Effect on the black hole shadow}

Electromagnetic waves with wavelengths longer than the plasma wavelength, $\lambda_{\rm p}=10.6\, (n_{\rm e}/10^7\,{\rm cm^{-3}})^{-1/2}\,{\rm m}$, will not penetrate through the plasma cloud towards the observer. The emission at shorter wavelengths is not blocked, but will experience scatter-broadening up to $\sim 1.4\,{\rm mm}$ \cite{1998ApJ...508L..61L} and hence any structure observed at this wavelength range is scatter-dominated and not source-dominated. Only at wavelengths $\lesssim 1.3\,{\rm mm}$, the radio structure of Sgr~A* starts to be source-dominated \cite{1998A&A...335L.106K,2008Natur.455...78D}. The enclosed curve on the sky plane that divides the region where photon geodesics intersect the event horizon from the region from where they can escape to infinity is of a particular importance. Its commonly known as the \textit{black hole shadow} and in principle should be resolved out by the current global mm-VLBI network \cite{2000ApJ...528L..13F} in case the light emission is not significantly scattered. To be more precise, the term \textit{silhouette} is more relevant in the context of our discussion. The term \textit{shadow} is related to a wall behind the black hole that is irradiated. The \textit{silhouette} is related to a radiating wall that is located behind the black hole. Considering the set-up of an accretion disk orbiting the black hole, the \textit{silhouette} is more appropriate. However, the term shadow is frequently used in general situations and we will apply it in the following discussion as well.

\begin{figure}[h!]
    \centering
    \includegraphics[width=\textwidth]{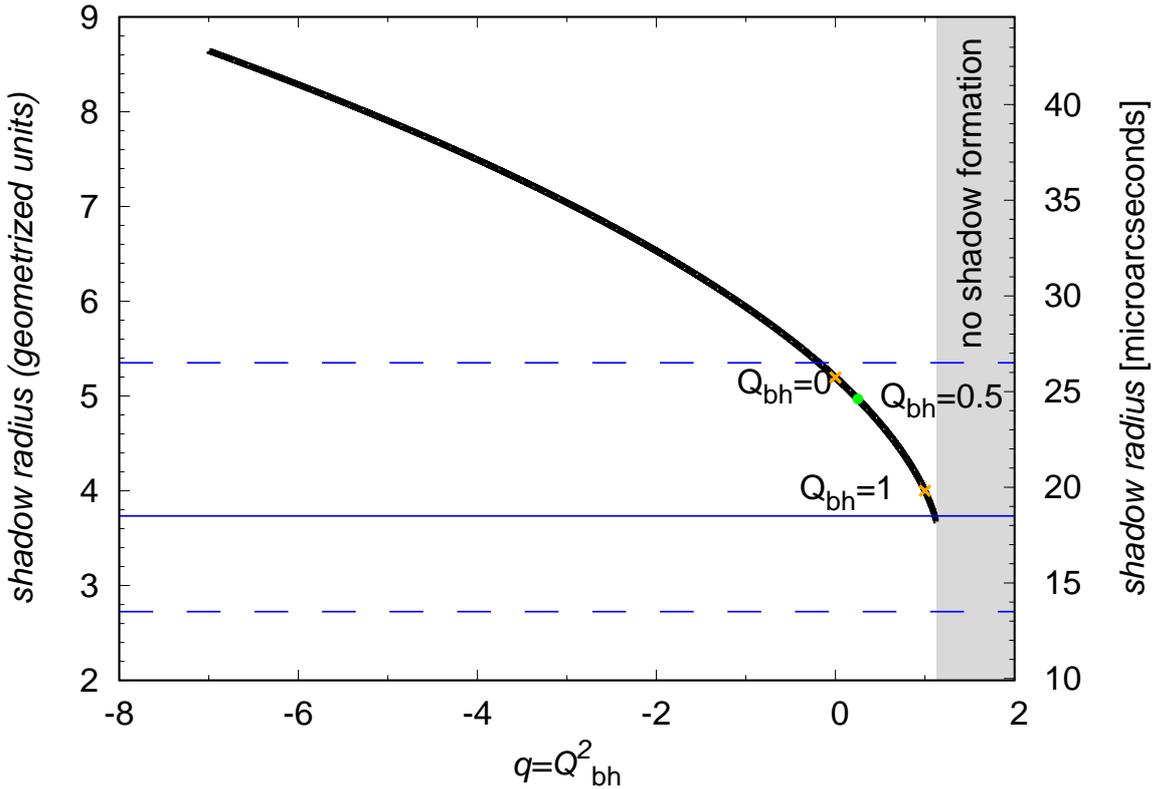}
    \caption{Black hole shadow size (radius) in geometrized units on the left axis as a function of parametrized charge, $q=Q_{\bullet}^2$, assuming the zero spin for this particular calculation. The right $y$-axis depicts the angular scale in microarcseconds for th distance and parameters of Sgr~A*. In addition, the orange crosses show the black-hole radius for an uncharged black hole, $Q_{\bullet}=0$, and for the extremal charge of $Q_{\bullet}=1$. The green point shows for an illustration the black hole charge of $Q_{\bullet}=0.5$. The shaded area depicts the charge values for which no shadow forms. The blue solid and dashed lines mark the VLBI core size range according to \cite{2008Natur.455...78D}.}
    \label{fig_black_hole_shadow}
\end{figure}

It is possible, in principle, to extract basic information about Sgr~A* by analyzing the black hole shadow size and shape provided that VLBI data are acquired and combined with high enough accuracy. Zakharov \cite{2014PhRvD..90f2007Z} proposed that the black hole shadow size could be used to discriminate between the charged and the uncharged case of Sgr~A*. They introduced parametrized charge $q=Q_{\bullet}^2$, which  encompasses both the normal positive electric charge in the Reissner-Nordstr\"om metric and the negative tidal charge. For the positive charge, the shadow size shrinks while for the negative tidal charge, it would increase, which was analysed by \cite{2014PhRvD..90f2007Z} and is depicted in Fig.~\ref{fig_black_hole_shadow} as a function of the parametrized charge $q$, assuming the zero spin (Reissner-Nordstr\"om metric). Using the VLBI measurements \cite{2008Natur.455...78D}, the VLBI core size of Sgr~A* was constrained to be $\theta_{\rm Sgr~A*}\approx 37^{+16}_{-10}\,{\rm \mu as}$, which seems to be more consistent with the shadow diameter of $8GM_{\bullet}/c^2\sim 39.4\,{\rm \mu as}$ corresponding to the black hole with an extremal electric charge when one considers the mean value only \cite{2014PhRvD..90f2007Z}. The shadow diameter for an uncharged Schwarzschild black hole is $6\sqrt{3}GM_{\bullet}/c^2\approx 51.2\,{\rm \mu as}$, i.e. it is by $\sim 30\%$ larger than for the extremal charge. However, within the uncertainties, the VLBI measurements are consistent with the whole range of possible electric charge values, $Q_{\bullet}=(0,1)$, see also Fig.~\ref{fig_black_hole_shadow}, and do not agree with the shadow sizes as predicted by the negative tidal charge.  

We also note that the VLBI core size is not related in an straightforward way to the shadow size. The brightest part of the accretion flow around Sgr~A* could be related to both the Doppler-boosted part of the flow and/or the footpoint of the jet \cite{2008Natur.455...78D,2009A&A...496...77F,2012MNRAS.421.1517D,2017FoPh...47..553E,2018ApJ...859...60L}. It can also be highly time-variable due to instabilities in the hot thick ADAF-type flow, where temporary \textit{hot spots} form and orbit the black hole on timescales of $45 \pm 15\,{\rm min}$ \cite{2018A&A...618L..10G}. Hence, the black hole shadow is not a clean variable and its size depends also on the black hole spin, which also decreases the shadow size. Moreover, the small value of electric charge analysed in the previous section, $Q_{\bullet}=(10^8,10^{15})\,{\rm C}$, cannot be detected via shadow measurements.

\subsection{Effect on the ISCO shift}

\begin{figure}
    \centering
    \includegraphics[width=0.49\textwidth]{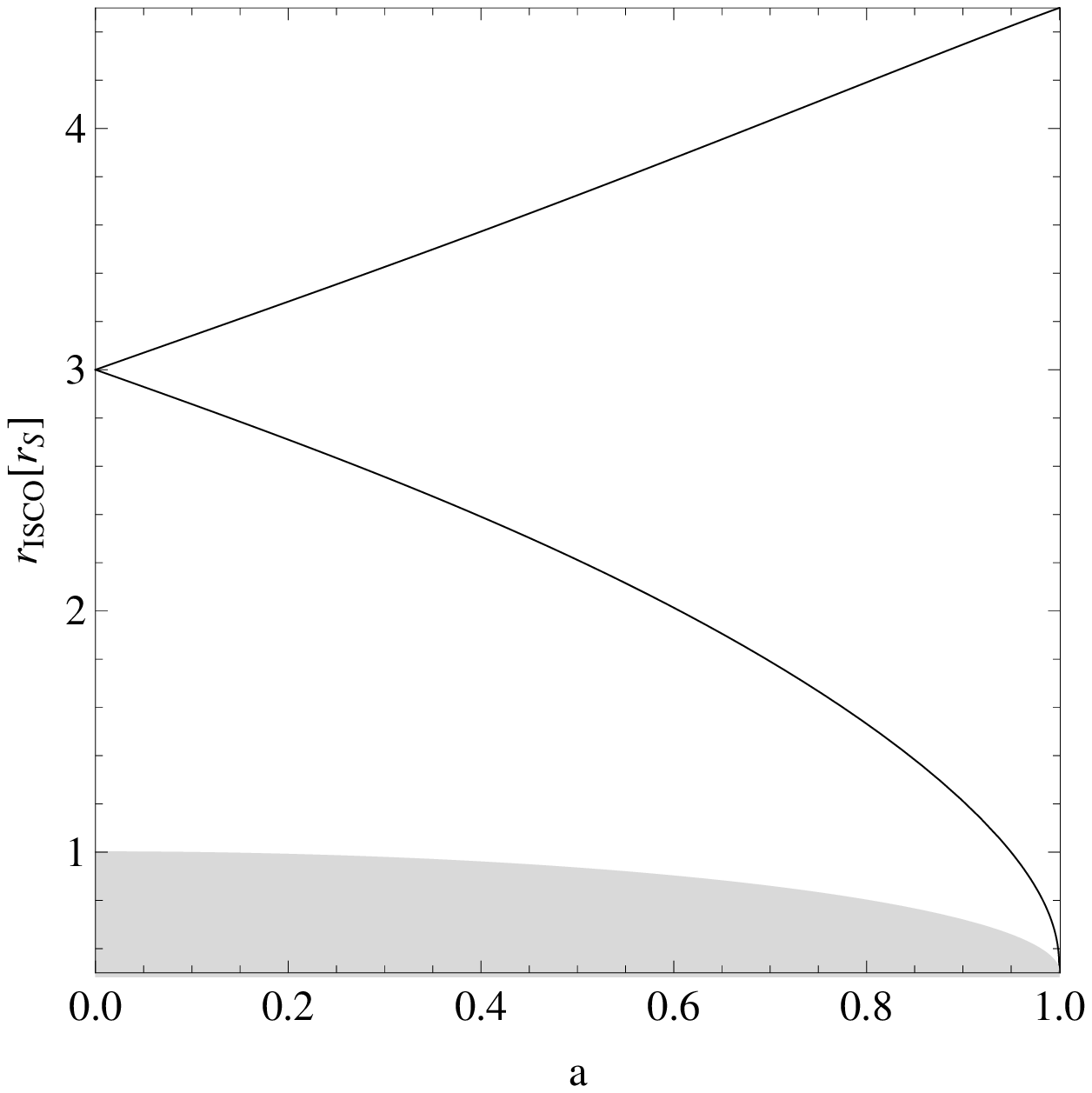}
    \includegraphics[width=0.48\textwidth]{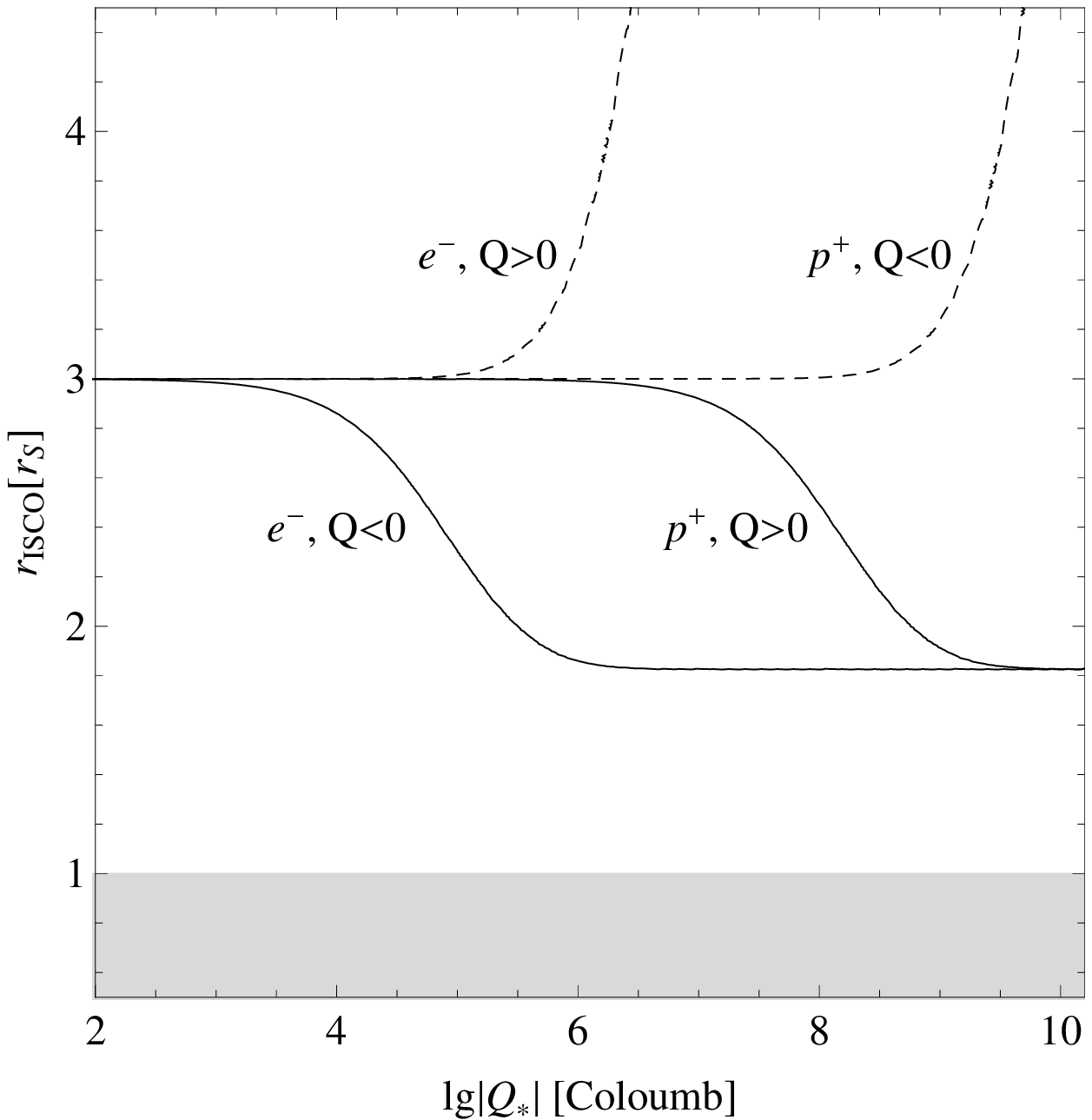}
    \caption{\textbf{Left panel:} The position of the ISCO as a function of the black hole spin: the upper line represents the retrograde spin, the lower line the prograde one. \textbf{Right panel:} The ISCO position for charged particles -- electrons ($e^{-}$) and protons ($p^{+}$) -- that orbit the SMBH (Sgr~A*) that has either positive $(Q_{\bullet}>0)$ or negative electric charge $(Q_{\bullet}<0)$.}
    \label{fig_ISCO}
\end{figure}

In an analogous way as the black hole spin, the charge shifts the innermost stable circular orbit (ISCO) of both neutral and charged particles \cite{2011PhRvD..83j4052P}. This has consequences for the dynamics of plasma in thick, hot ADAF-type flows present in Sgr~A* system. In addition, it introduces a degeneracy in terms of reliable spin determination, which for Sgr~A* has been based on the periodicity and the related ISCO location. For the orbital period of prograde-orbiting particles in the equatorial plane, the following period relation applies, $P=2\pi (r_{\rm ISCO}^{3/2}+a_{\bullet})GM_{\bullet}/c^3$. In Fig.~\ref{fig_ISCO} (left panel), we plot the dependency of the ISCO on the black-hole spin for a Kerr black hole (the upper line is for retrograde spin, the lower line is for prograde spin). 

Even for the small charge values of Sgr~A* summarized in the previous section, the ISCO shift is profound for charged particles (electrons and protons). In Fig.~\ref{fig_ISCO} (right panel), we plot the dependency of the ISCO location for free charged particles that orbit the non-rotating SMBH with a small charge. The ISCO shifts from $r_{\rm ISCO}=3r_{\rm s}$, which applies to an uncharged black hole, to smaller or larger radii for relatively small black-hole charges of $Q_{\bullet}=10^4-10^5\,{\rm C}$. For like charges, ($e^{-}$, $Q_{\bullet}<0$ and $p^{+}$, $Q_{\bullet}>0$), the ISCO can shift up to $r_{\rm ISCO}=1.83r_{\rm s}$, which effectively mimics the prograde spin of $a_{\bullet}=0.64$. This effect should be taken into account, as this falls into the range of the inferred spin of Sgr~A*, $a_{\bullet}\gtrsim 0.4$ \cite{2006A&A...460...15M,2010MNRAS.403L..74K,2010A&A...510A...3Z,2018ApJ...863...15W}. The ISCO shift in turn affects the Lorentz factors of orbiting particles that emit synchrotron and/or contribute to the inverse Compton effect close to the ISCO of Sgr~A*. In addition, the gravitational redshift $z$ increases from $z=0.225$ to $z=0485$ for the corresponding ISCO shift of $r_{\rm ISCO}=3r_{\rm s}$ to $r_{\rm ISCO}=1.83r_{\rm s}$.    

In addition to the influence of the electric charge of the SMBH, an external magnetic field can also have a profound influence on the ISCO; see e.g. \cite{2010PhRvD..82h4034F,2014arXiv1410.1663P}. Hence, the combined effect of the black hole charge and that of the surrounding electromagnetic field introduces an extra uncertainty into the spin determination of astrophysical black holes in general.   

\subsection{Effect on the thermal bremsstrahlung profile}

\begin{figure}[h!]
    \centering
    \includegraphics[width=0.48\textwidth]{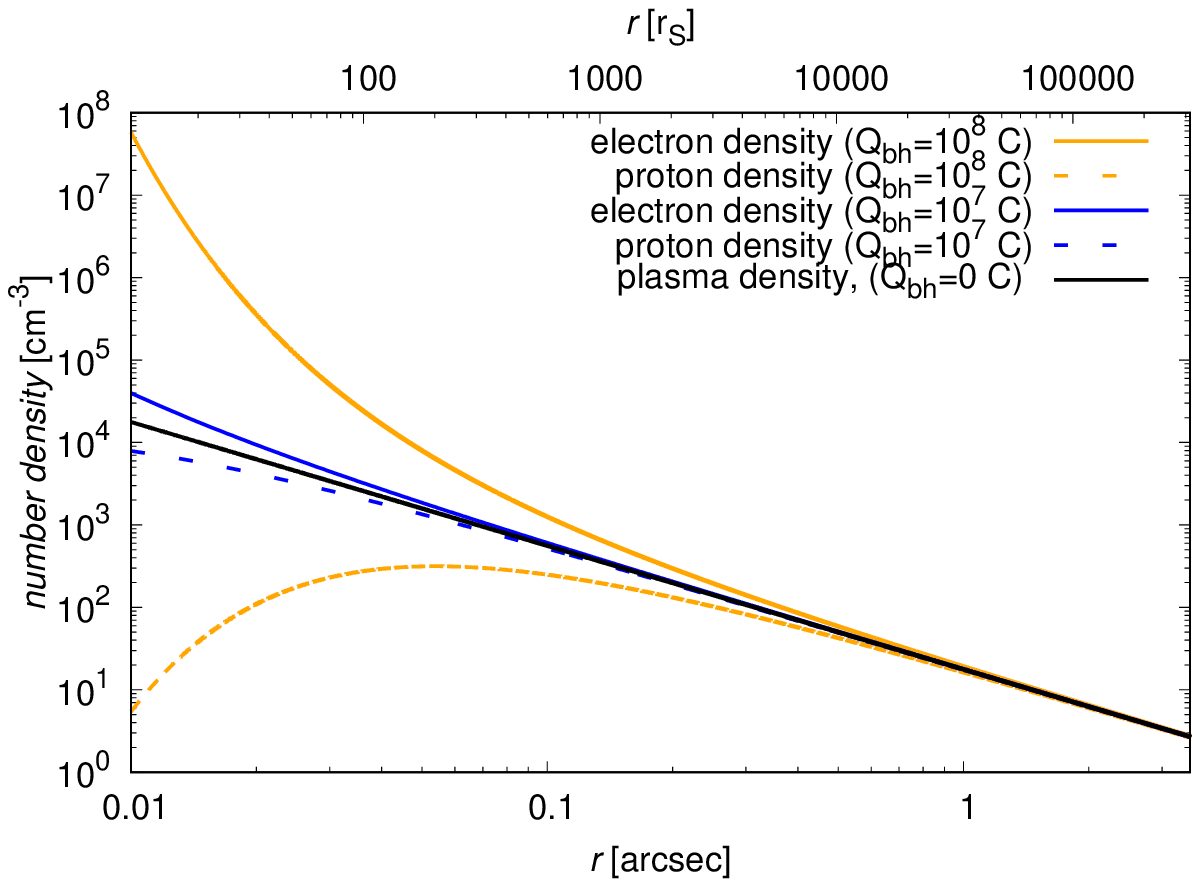} 
    \includegraphics[width=0.48\textwidth]{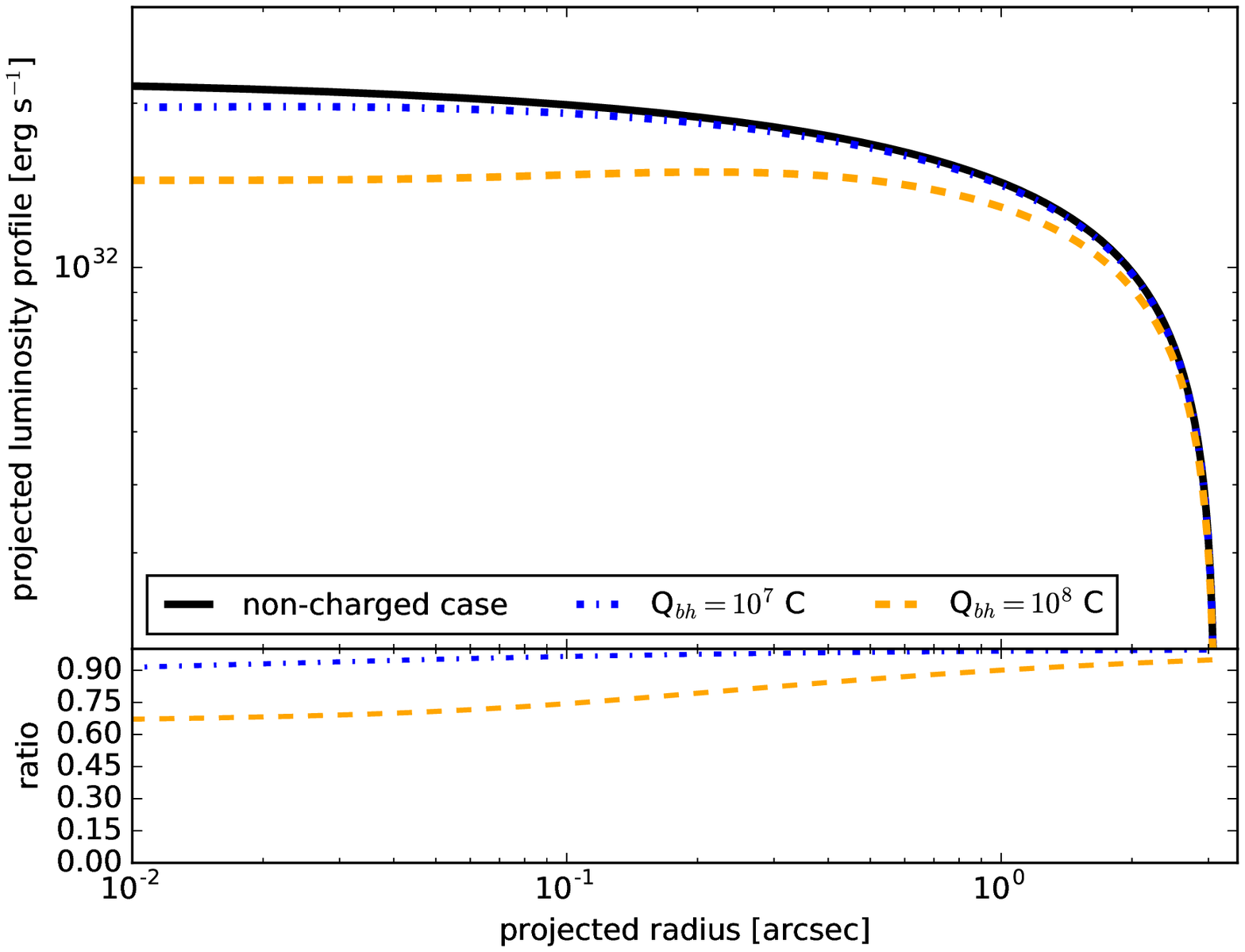}
    \caption{\textbf{Left panel:} Number density profile of electrons and protons in hot diluted plasma around Sgr~A* with the charge of $0$, $10^7$, and $10^8$ Coulombs. \textbf{Right panel:} The corresponding projected surface brightness of thermal bremsstrahlung calculated for the same electric charges and on the same length-scales as the left panel. For details, see \cite{2018MNRAS.480.4408Z}.}
    \label{fig_brems}
\end{figure}

In \cite{2018MNRAS.480.4408Z}, we propose a test of the presence of the electric charge associated with Sgr~A* by looking at the projected flux density profile of the thermal X-ray bremsstrahlung. The essence of the test lies in the assumption that an unshielded electric charge would cause the charge separation in the vicinity of Sgr~A*, e.g. a positively charged black hole would cause, in a stationary set-up, the electron number density to increase towards the centre, while the proton number density would exponentially decrease and vice versa for a negatively charged black hole. Although the assumption of stationarity may seem as an oversimplification, it can still serve as a basis for calculating the synthetic projected surface brightness of the thermal bremsstrahlung that is subsequently compared to the observed one, as determined by e.g. \cite{2015A&A...581A..64R}. In Fig.~\ref{fig_brems} (left panel), we show the number density of protons and electrons for the positive charges of $10^7$ and $10^8$ Coulombs as exemplary values. In the right panel of Fig.~\ref{fig_brems}, the corresponding surface brightness profiles of the thermal bremsstrahlung are shown (for details, see \cite{2018MNRAS.480.4408Z}). The non-zero charge of Sgr~A* leads to the charge separation -- the number density of free electrons increases towards the centre, while the number density of protons decreases. In terms of the bremsstrahlung emission, scattering of like particles ($e-e$ and $p-p$) is much less efficient than electron-proton scattering. This is translated in the surface brightness profile, which for the non-zero charge of Sgr~A* flattens and eventually drops and decreases for smaller radii with respect to the non-charge case, see Fig.~\ref{fig_brems} (right panel). The observed bremsstrahlung brightness profile, see \cite{2015A&A...581A..64R}, is consistent with a slightly rising to flat profile, which puts an upper limit on the positive electric charge of Sgr~A*, $Q_{\bullet, Sgr~A*}\lesssim 3 \times 10^8\,{\rm C}$.         

\subsection{Effect on the energy extraction mechanism}
According to the black hole thermodynamics, up to $29\%$ of the total energy of rotating black holes is available for the  extraction~\cite{Bar-Car-Haw:1973:CMP:}. In general, the possibility of existence of negative energy states inside the ergosphere of rotating black hole may result in extraction of its rotational energy. Among astrophysically relevant energy extraction mechanisms one can distinguish essentially two leading ones, namely, Blandford-Znajek mechanism (BZ)~\cite{1977MNRAS.179..433B} and magnetic Penrose process (MPP)~\cite{1985ApJ...290...12W}. The former is generally accepted as the leading mechanism for production of relativistic black hole jets supported by various numerical simulations. Both of these processes require a rotating black hole and the presence of external magnetic field, which makes two processes similar, although the processes operate with different efficiencies~\cite{2018MNRAS.478L..89D}. 
In the presence of black hole charge (which can be produced by twisting of magnetic field lines due to the rotation of a black hole), Coloumbic interaction of matter with black hole gives an additional contribution to the negative energy inflow, thereby converting the rotational energy of the black hole into extractable electromagnetic energy. A careful look into equations constituting MPP and BZ shows that the driving engine of the both lies in the existence of spin-induced electric field due to frame-dragging effect on magnetic field lines. Discharge of spin-induced black hole charge by negative energy flux causes the decrease of the spin of the black hole and resulting extraction of black hole's rotational energy. Thus, the presence of black hole charge may support an acceleration of charged particles to relativistic velocities as seen e.g. in the relativistic jets of black holes and can be relevant also in other similar high-energy phenomena.

\section{Summary}

We used the observational data on the immediate black hole surrounding to
constrain the electric charge of Sgr A*, which is the third parameter
for black holes according to the no-hair theorem. The detailed analysis is presented in \cite{2018MNRAS.480.4408Z}. The main findings may be summarized as follows:
\begin{itemize}
  \item previous claims that astrophysical black holes can be treated as nearly uncharged, with $Q_{\bullet}\lesssim 10^{-18}\,Q_{\rm max}^{\rm norot}$ \cite{1975ARA&A..13..381E}, is not supported for the Galactic centre black hole and potentially other astrophysical black hole systems, which is in agreement with other studies, e.g. concerning black hole--neutron star binaries and associated merger events \cite{PhysRevD.98.123002},
  \item however,  the black hole charge is small, with the potential upper theoretical limit of $Q_{\rm \bullet ind}^{\rm max}<10^{15}\,{\rm C}$ (twelve orders of magnitude below the extremal value), hence the space-time metric is not significantly affected, i.e. the motion of neutral bodies can be analysed in the framework of Kerr metric in most cases, 
  \item on the other hand, a certain degree of caution is always at place, since even such a small charge can substantially affect the dynamics of charged particles (shift of ISCO; see \cite{2017PhRvD..96f3015S,2018MNRAS.480.4408Z}) and is related essentially to energy extraction from black holes (Blandford-Znajek process or Magnetic Penrose process),
  \item we found that a small positive unshielded charge can have an impact on the electron/proton density distribution in the surrounding plasma, on which the thermal bremsstrahlung emissivity depends, $L_{\rm brems} \propto Z^2n_{\rm i} n_{\rm e}$. Based on that, we propose a novel observational test of the black hole charge presence based on the flattening and eventually drop in the X-ray bremsstrahlung profile for increasing values of a (positive) electric charge. The current X-ray data are consistent with the flat to slightly rising profile (not decreasing), which provides an observational limit on the charge of Sgr~A*, $Q_{\rm \bullet Sgr A^{*}}\lesssim 3\times 10^{8}\,{\rm C}$.
\end{itemize}

In conclusion, we showed that a small charge that is negligible for space-time structure may be of relevance for the plasma dynamics close to the Galactic centre black hole and in an analogous way for other supermassive black holes. The relevance of charge and the associated electromagnetic signal may be even greater in rare, but relevant astrophysical situations -- plunges of neutron stars into supermassive black holes \cite{2017CoSka..47..124K,2018PhRvD..98h4055K} or black hole-neutron star mergers \cite{PhysRevD.98.123002}, when the strong magnetic field associated with the neutron star threads the black-hole horizon and the induced stable electric charge can reach large values proportional to the magnetic-field strength according to Eq.~\eqref{eq_max_induced_charge}.

\section*{Acknowledgements}

We thank the organizers of the FISICPAC-2018 conference at the University of Sharjah for the overall organization of a very nice event. The presenting author Michal Zaja\v{c}ek acknowledges the financial support from the National Science Centre, Poland, grant No. 2017/26/A/ST9/00756 (Maestro 9). 
Arman Tursunov was supported by International Mobility Project CZ.02.2.69/0.0/0.0/16\_027/0008521.

\bibliographystyle{iopart-num}
\bibliography{zajacek}

\providecommand{\newblock}{}
\begin{thebibliography}{10}
\expandafter\ifx\csname url\endcsname\relax
  \def\url#1{{\tt #1}}\fi
\expandafter\ifx\csname urlprefix\endcsname\relax\def\urlprefix{URL }\fi
\providecommand{\eprint}[2][]{\url{#2}}

\bibitem{1974ApJ...194..265B}
{Balick} B and {Brown} R~L 1974 {\em \apj\/} {\bf 194} 265--270

\bibitem{1971Natur.233..112D}
{Downes} D and {Martin} A~H~M 1971 {\em \nat\/} {\bf 233} 112--114

\bibitem{1973ApJ...184..415R}
{Rieke} G~H and {Low} F~J 1973 {\em \apj\/} {\bf 184} 415--425

\bibitem{1974ApJ...192..325B}
{Balick} B and {Sanders} R~H 1974 {\em \apj\/} {\bf 192} 325--336

\bibitem{1969Natur.223..690L}
{Lynden-Bell} D 1969 {\em \nat\/} {\bf 223} 690--694

\bibitem{1971MNRAS.152..461L}
{Lynden-Bell} D and {Rees} M~J 1971 {\em \mnras\/} {\bf 152} 461

\bibitem{1975ApJ...202L..63L}
{Lo} K~Y, {Schilizzi} R~T, {Cohen} M~H and {Ross} H~N 1975 {\em \apjl\/} {\bf
  202} L63--L65

\bibitem{1981ApJ...250..155B}
{Brown} R~L, {Johnston} K~J and {Lo} K~Y 1981 {\em \apj\/} {\bf 250} 155--159

\bibitem{1978ApJ...219..121B}
{Becklin} E~E, {Matthews} K, {Neugebauer} G and {Willner} S~P 1978 {\em \apj\/}
  {\bf 219} 121--128

\bibitem{1978ApJ...220..831B}
{Becklin} E~E, {Matthews} K, {Neugebauer} G and {Willner} S~P 1978 {\em \apj\/}
  {\bf 220} 831--835

\bibitem{1976ApJ...205L...5W}
{Wollman} E~R, {Geballe} T~R, {Lacy} J~H, {Townes} C~H and {Rank} D~M 1976 {\em
  \apjl\/} {\bf 205} L5--L9

\bibitem{1977ApJ...218L.103W}
{Wollman} E~R, {Geballe} T~R, {Lacy} J~H, {Townes} C~H and {Rank} D~M 1977 {\em
  \apjl\/} {\bf 218} L103--L107

\bibitem{1979ApJ...227L..17L}
{Lacy} J~H, {Baas} F, {Townes} C~H and {Geballe} T~R 1979 {\em \apjl\/} {\bf
  227} L17--L20

\bibitem{1980ApJ...241..132L}
{Lacy} J~H, {Townes} C~H, {Geballe} T~R and {Hollenbach} D~J 1980 {\em \apj\/}
  {\bf 241} 132--146

\bibitem{1982ApJ...262..110B}
{Brown} R~L 1982 {\em \apj\/} {\bf 262} 110--119

\bibitem{1984ApJ...276..551G}
{Genzel} R, {Watson} D~M, {Townes} C~H, {Dinerstein} H~L, {Hollenbach} D,
  {Lester} D~F, {Werner} M and {Storey} J~W~V 1984 {\em \apj\/} {\bf 276}
  551--559

\bibitem{1982ApJ...262..120L}
{Lacy} J~H, {Townes} C~H and {Hollenbach} D~J 1982 {\em \apj\/} {\bf 262}
  120--134

\bibitem{1983Natur.301..661T}
{Townes} C~H, {Lacy} J~H, {Geballe} T~R and {Hollenbach} D~J 1983 {\em \nat\/}
  {\bf 301} 661--666

\bibitem{1982ApJ...253..108B}
{Brown} R~L and {Lo} K~Y 1982 {\em \apj\/} {\bf 253} 108--114

\bibitem{2018A&A...615L..15G_new}
{Gravity Collaboration} 2018 {\em \aap\/} {\bf 615} L15 (\textit{Preprint}
  \eprint{1807.09409})

\bibitem{2018A&A...618L..10G}
{Gravity Collaboration} 2018 {\em \aap\/} {\bf 618} L10 (\textit{Preprint}
  \eprint{1810.12641})

\bibitem{2005MNRAS.363..353B}
{Broderick} A~E and {Loeb} A 2005 {\em \mnras\/} {\bf 363} 353--362
  (\textit{Preprint} \eprint{astro-ph/0506433})

\bibitem{2006ApJ...636L.109B}
{Broderick} A~E and {Loeb} A 2006 {\em \apjl\/} {\bf 636} L109--L112
  (\textit{Preprint} \eprint{astro-ph/0508386})

\bibitem{2006A&A...460...15M}
{Meyer} L, {Eckart} A, {Sch{\"o}del} R, {Duschl} W~J, {Mu{\v z}i{\'c}} K,
  {Dov{\v c}iak} M and {Karas} V 2006 {\em \aap\/} {\bf 460} 15--21
  (\textit{Preprint} \eprint{astro-ph/0610104})

\bibitem{2017MNRAS.472.4422K}
{Karssen} G~D, {Bursa} M, {Eckart} A, {Valencia-S} M, {Dov{\v c}iak} M, {Karas}
  V and {Hor{\'a}k} J 2017 {\em \mnras\/} {\bf 472} 4422--4433
  (\textit{Preprint} \eprint{1709.09896})

\bibitem{2008poii.conf..431E}
{Eisenhauer} F, {Perrin} G, {Rabien} S, {Eckart} A, {L{\'e}na} P, {Genzel} R,
  {Abuter} R, {Paumard} T and {Brandner} W 2008 {\em The Power of Optical/IR
  Interferometry: Recent Scientific Results and 2nd Generation\/} ed {Richichi}
  A, {Delplancke} F, {Paresce} F and {Chelli} A p 431

\bibitem{2019hsax.conf..609A}
{Abuter} R, {Amorim} A, {Anugu} N, {Baub{\"o}ck} M, {Benisty} M, {Berger} J~P,
  {Blind} N, {Bonnet} H, {Brandner} W, {Buron} A, {Collin} C, {Chapron} F,
  {Cl{\'e}net} Y, {Coud{\'e} du Foresto} V, {de Zeeuw} P~T, {Deen} C,
  {Delplancke-Str{\"o}bele} F, {Dembet} R, {Dexter} J, {Duvert} G, {Eckart} A,
  {Eisenhauer} F, {Finger} G, {F{\"o}rster Schreiber} N~M, {F{\'e}dou} P,
  {Garcia} P, {Garcia Lopez} R, {Gao} F, {Gendron} E, {Genzel} R, {Gillessen}
  S, {Gordo} P, {Habibi} M, {Haubois} X, {Haug} M, {Hau{\ss}mann} F, {Henning}
  T, {Hippler} S, {Horrobin} M, {Hubert} Z, {Hubin} N, {Jimenez Rosales} A,
  {Jochum} L, {Jocou} L, {Kaufer} A, {Kellner} S, {Kendrew} S, {Kervella} P,
  {Kok} Y, {Kulas} M, {Lacour} S, {Lapeyr{\`e}re} V, {Lazareff} V, {Le Bouquin}
  J~B, {L{\'e}na} P, {Lippa} M, {Lenzen} R, {M{\'e}rand} A, {M{\"u}ler} E,
  {Neumann} U, {Ott} T, {Palanca} L, {Paumard} T, {Pasquini} L, {Perraut} K,
  {Perrin} G, {Pfuhl} O, {Plewa} P~M, {Rabien} S, {Ram{\'{\i}}rez} A, {Ramos}
  J, {Rau} C, {Rodr{\'{\i}}guez-Coira} G, {Rohloff} R~R, {Rousset} G,
  {Sanchez-Bermudez} J, {Scheithauer} S, {Sch{\"o}ller} M, {Schuler} N,
  {Spyromilio} J, {Straub} O, {Straubmeier} C, {Sturm} E, {Tacconi} L~J,
  {Tristram} K~R~W, {Vincent} F, {von Fellenberg} S, {Wank} I, {Waisberg} I,
  {Widmann} F, {Wieprecht} Wiest M~F, {Wiezorrek} E, {Woillez} J, {Yazici} S,
  {Ziegler} S and {Zins} G 2019 {\em Highlights on Spanish Astrophysics X,
  Proceedings of the XIII Scientific Meeting of the Spanish Astronomical
  Society held on July 16-20, 2018, in Salamanca, Spain, ISBN
  978-84-09-09331-1. B. Montesinos, A. Asensio Ramos, F. Buitrago, R.
  Sch{\"o}del, E. Villaver, S. P{\'e}rez-Hoyos, I. Ord{\'o}{\~n}ez-Etxeberria
  (eds.) p. 609-610\/} ed {Montesinos} B, {Asensio Ramos} A, {Buitrago} F,
  {Sch{\"o}del} R, {Villaver} E, {P{\'e}rez-Hoyos} S and
  {Ord{\'o}{\~n}ez-Etxeberria} I pp 609--610

\bibitem{2018ApJ...859...60L}
{Lu} R~S, {Krichbaum} T~P, {Roy} A~L, {Fish} V~L, {Doeleman} S~S, {Johnson}
  M~D, {Akiyama} K, {Psaltis} D, {Alef} W, {Asada} K, {Beaudoin} C, {Bertarini}
  A, {Blackburn} L, {Blundell} R, {Bower} G~C, {Brinkerink} C, {Broderick} A~E,
  {Cappallo} R, {Crew} G~B, {Dexter} J, {Dexter} M, {Falcke} H, {Freund} R,
  {Friberg} P, {Greer} C~H, {Gurwell} M~A, {Ho} P~T~P, {Honma} M, {Inoue} M,
  {Kim} J, {Lamb} J, {Lindqvist} M, {Macmahon} D, {Marrone} D~P,
  {Mart{\'{\i}}-Vidal} I, {Menten} K~M, {Moran} J~M, {Nagar} N~M, {Plambeck}
  R~L, {Primiani} R~A, {Rogers} A~E~E, {Ros} E, {Rottmann} H, {SooHoo} J,
  {Spilker} J, {Stone} J, {Strittmatter} P, {Tilanus} R~P~J, {Titus} M,
  {Vertatschitsch} L, {Wagner} J, {Weintroub} J, {Wright} M, {Young} K~H,
  {Zensus} J~A and {Ziurys} L~M 2018 {\em \apj\/} {\bf 859} 60
  (\textit{Preprint} \eprint{1805.09223})

\bibitem{2017ApJ...845...22P}
{Parsa} M, {Eckart} A, {Shahzamanian} B, {Karas} V, {Zaja{\v c}ek} M, {Zensus}
  J~A and {Straubmeier} C 2017 {\em \apj\/} {\bf 845} 22 (\textit{Preprint}
  \eprint{1708.03507})

\bibitem{2017FoPh...47..553E}
{Eckart} A, {H{\"u}ttemann} A, {Kiefer} C, {Britzen} S, {Zaja{\v c}ek} M,
  {L{\"a}mmerzahl} C, {St{\"o}ckler} M, {Valencia-S} M, {Karas} V and
  {Garc{\'{\i}}a-Mar{\'{\i}}n} M 2017 {\em Foundations of Physics\/} {\bf 47}
  553--624 (\textit{Preprint} \eprint{1703.09118})

\bibitem{1998ApJ...492..554N}
{Narayan} R, {Mahadevan} R, {Grindlay} J~E, {Popham} R~G and {Gammie} C 1998
  {\em \apj\/} {\bf 492} 554--568 (\textit{Preprint} \eprint{astro-ph/9706112})

\bibitem{2002A&A...396L..31A}
{Abramowicz} M~A, {Klu{\'z}niak} W and {Lasota} J~P 2002 {\em \aap\/} {\bf 396}
  L31--L34 (\textit{Preprint} \eprint{astro-ph/0207270})

\bibitem{2016PhRvL.116q1101C}
{Cardoso} V, {Franzin} E and {Pani} P 2016 {\em Physical Review Letters\/} {\bf
  116} 171101 (\textit{Preprint} \eprint{1602.07309})

\bibitem{2007CQGra..24.4191C}
{Chirenti} C~B~M~H and {Rezzolla} L 2007 {\em Classical and Quantum Gravity\/}
  {\bf 24} 4191--4206 (\textit{Preprint} \eprint{0706.1513})

\bibitem{2016PhRvD..94h4016C}
{Chirenti} C and {Rezzolla} L 2016 {\em \prd\/} {\bf 94} 084016
  (\textit{Preprint} \eprint{1602.08759})

\bibitem{Bardeen1968}
Bardeen J~M 1968 {\em Proceedings of International Conference GR5, Tiflis,
  U.S.S.R.\/} p 174

\bibitem{Ayon-Beato1998}
{Ay{\'o}n-Beato} E and {Garc{\'\i}a} A 1998 {\em \prl\/} {\bf 80} 5056--5059
  (\textit{Preprint} \eprint{gr-qc/9911046})

\bibitem{PhysRevD.76.064024}
Werner M~C and Petters A~O 2007 {\em Phys. Rev. D\/} {\bf 76}(6) 064024
  \urlprefix\url{https://link.aps.org/doi/10.1103/PhysRevD.76.064024}

\bibitem{2003Natur.425..934G}
{Genzel} R, {Sch{\"o}del} R, {Ott} T, {Eckart} A, {Alexander} T, {Lacombe} F,
  {Rouan} D and {Aschenbach} B 2003 {\em \nat\/} {\bf 425} 934--937
  (\textit{Preprint} \eprint{astro-ph/0310821})

\bibitem{2012A&A...537A..52E}
{Eckart} A, {Garc{\'{\i}}a-Mar{\'{\i}}n} M, {Vogel} S~N, {Teuben} P, {Morris}
  M~R, {Baganoff} F, {Dexter} J, {Sch{\"o}del} R, {Witzel} G, {Valencia-S} M,
  {Karas} V, {Kunneriath} D, {Straubmeier} C, {Moser} L, {Sabha} N, {Buchholz}
  R, {Zamaninasab} M, {Mu{\v z}i{\'c}} K, {Moultaka} J and {Zensus} J~A 2012
  {\em \aap\/} {\bf 537} A52

\bibitem{2018ApJ...863...15W}
{Witzel} G, {Martinez} G, {Hora} J, {Willner} S~P, {Morris} M~R, {Gammie} C,
  {Becklin} E~E, {Ashby} M~L~N, {Baganoff} F, {Carey} S, {Do} T, {Fazio} G~G,
  {Ghez} A, {Glaccum} W~J, {Haggard} D, {Herrero-Illana} R, {Ingalls} J,
  {Narayan} R and {Smith} H~A 2018 {\em \apj\/} {\bf 863} 15 (\textit{Preprint}
  \eprint{1806.00479})

\bibitem{2003bhcg.book.....M}
{Melia} F 2003 {\em {The black hole at the center of our galaxy (Princeton, NJ:
  Princeton University Press)}\/}

\bibitem{2018MNRAS.480.4408Z}
{Zaja{\v c}ek} M, {Tursunov} A, {Eckart} A and {Britzen} S 2018 {\em \mnras\/}
  {\bf 480} 4408--4423 (\textit{Preprint} \eprint{1808.07327})

\bibitem{1974PhRvD..10.1680W}
{Wald} R~M 1974 {\em \prd\/} {\bf 10} 1680--1685

\bibitem{2014ApJ...787..117K}
{Kop{\'a}{\v c}ek} O and {Karas} V 2014 {\em \apj\/} {\bf 787} 117
  (\textit{Preprint} \eprint{1404.5495})

\bibitem{kovar2010}
Kovar J, Kopacek O, Karas V and Stuchlik Z {2010} {\em {CLASSICAL AND QUANTUM
  GRAVITY}\/} {\bf {27}} ISSN {0264-9381}

\bibitem{kopacek2010}
Kopacek O, Karas V, Kovar J and Stuchlik Z {2010} {\em {ASTROPHYSICAL
  JOURNAL}\/} {\bf {722}} {1240--1259} ISSN {0004-637X}

\bibitem{2015CQGra..32p5009K}
{Kolo{\v s}} M, {Stuchl{\'{\i}}k} Z and {Tursunov} A 2015 {\em Classical and
  Quantum Gravity\/} {\bf 32} 165009 (\textit{Preprint} \eprint{1506.06799})

\bibitem{2016PhRvD..93h4012T}
{Tursunov} A, {Stuchl{\'{\i}}k} Z and {Kolo{\v s}} M 2016 {\em \prd\/} {\bf 93}
  084012 (\textit{Preprint} \eprint{1603.07264})

\bibitem{Stuchlik2016}
Stuchl{\'i}k Z and Kolo{\v{s}} M 2016 {\em The European Physical Journal C\/}
  {\bf 76} 32 ISSN 1434-6052
  \urlprefix\url{https://doi.org/10.1140/epjc/s10052-015-3862-2}

\bibitem{Kolos2017}
Kolo{\v{s}} M, Tursunov A and Stuchl{\'i}k Z 2017 {\em The European Physical
  Journal C\/} {\bf 77} 860 ISSN 1434-6052
  \urlprefix\url{https://doi.org/10.1140/epjc/s10052-017-5431-3}

\bibitem{2018ApJ...861....2T}
{Tursunov} A, {Kolo{\v s}} M, {Stuchl{\'{\i}}k} Z and {Gal'tsov} D~V 2018 {\em
  \apj\/} {\bf 861} 2 (\textit{Preprint} \eprint{1803.09682})

\bibitem{1991JPhy1...1.1005K}
{Karas} V and {Vokrouhlick{\'y}} D 1991 {\em Journal de Physique I\/} {\bf 1}
  1005--1012

\bibitem{1991JMP....32..714K}
{Karas} V and {Vokrouhlick{\'y}} D 1991 {\em Journal of Mathematical Physics\/}
  {\bf 32} 714--716

\bibitem{2011PhRvD..84h4002K}
{Kov{\'a}{\v r}} J, {Slan{\'y}} P, {Stuchl{\'{\i}}k} Z, {Karas} V,
  {Cremaschini} C and {Miller} J~C 2011 {\em \prd\/} {\bf 84} 084002
  (\textit{Preprint} \eprint{1110.4843})

\bibitem{2011PhRvD..83j4052P}
{Pugliese} D, {Quevedo} H and {Ruffini} R 2011 {\em \prd\/} {\bf 83} 104052
  (\textit{Preprint} \eprint{1103.1807})

\bibitem{2017PhRvD..96f3015S}
{Schroven} K, {Hackmann} E and {L{\"a}mmerzahl} C 2017 {\em \prd\/} {\bf 96}
  063015

\bibitem{2018PhRvD..97j4019T}
{Trova} A, {Schroven} K, {Hackmann} E, {Karas} V, {Kov{\'a}{\v{r}}} J and
  {Slan{\'y}} P 2018 {\em \prd\/} {\bf 97} 104019

\bibitem{slany2013}
Slany P, Kovar J, Stuchlik Z and Karas V {2013} {\em {ASTROPHYSICAL JOURNAL
  SUPPLEMENT SERIES}\/} {\bf {205}} ISSN {0067-0049}

\bibitem{cremaschini2013}
Cremaschini C, Kovar J, Slany P, Stuchlik Z and Karas V {2013} {\em
  {ASTROPHYSICAL JOURNAL SUPPLEMENT SERIES}\/} {\bf {209}} ISSN {0067-0049}

\bibitem{kovar2014}
Kovar J, Slany P, Cremaschini C, Stuchlik Z, Karas V and Trova A {2014} {\em
  {PHYSICAL REVIEW D}\/} {\bf {90}} ISSN {1550-7998}

\bibitem{2012GReGr..44.1753I}
{Iorio} L 2012 {\em General Relativity and Gravitation\/} {\bf 44} 1753--1767
  (\textit{Preprint} \eprint{1112.3520})

\bibitem{2014PhRvD..90f2007Z}
{Zakharov} A~F 2014 {\em \prd\/} {\bf 90} 062007 (\textit{Preprint}
  \eprint{1407.7457})

\bibitem{2013Sci...341..981W}
{Wang} Q~D, {Nowak} M~A, {Markoff} S~B, {Baganoff} F~K, {Nayakshin} S, {Yuan}
  F, {Cuadra} J, {Davis} J, {Dexter} J, {Fabian} A~C, {Grosso} N, {Haggard} D,
  {Houck} J, {Ji} L, {Li} Z, {Neilsen} J, {Porquet} D, {Ripple} F and
  {Shcherbakov} R~V 2013 {\em Science\/} {\bf 341} 981--983 (\textit{Preprint}
  \eprint{1307.5845})

\bibitem{2003ApJ...591..891B}
{Baganoff} F~K, {Maeda} Y, {Morris} M, {Bautz} M~W, {Brandt} W~N, {Cui} W,
  {Doty} J~P, {Feigelson} E~D, {Garmire} G~P, {Pravdo} S~H, {Ricker} G~R and
  {Townsley} L~K 2003 {\em \apj\/} {\bf 591} 891--915 (\textit{Preprint}
  \eprint{astro-ph/0102151})

\bibitem{2015A&A...581A..64R}
{R{\'o}{\.z}a{\'n}ska} A, {Mr{\'o}z} P, {Mo{\'s}cibrodzka} M, {Sobolewska} M
  and {Adhikari} T~P 2015 {\em \aap\/} {\bf 581} A64 (\textit{Preprint}
  \eprint{1507.01798})

\bibitem{1926ics..book.....E}
{Eddington} A~S 1926 {\em {The Internal Constitution of the Stars (Cambridge:
  Cambridge University Press)}\/}

\bibitem{2001A&A...372..913N}
{Neslu{\v s}an} L 2001 {\em \aap\/} {\bf 372} 913--915

\bibitem{1978ApJ...220..743B}
{Bally} J and {Harrison} E~R 1978 {\em \apj\/} {\bf 220} 743

\bibitem{1963PhRvL..11..237K}
{Kerr} R~P 1963 {\em Physical Review Letters\/} {\bf 11} 237--238

\bibitem{1965JMP.....6..918N}
{Newman} E~T, {Couch} E, {Chinnapared} K, {Exton} A, {Prakash} A and {Torrence}
  R 1965 {\em Journal of Mathematical Physics\/} {\bf 6} 918--919

\bibitem{2010RvMP...82.3121G}
{Genzel} R, {Eisenhauer} F and {Gillessen} S 2010 {\em Reviews of Modern
  Physics\/} {\bf 82} 3121--3195 (\textit{Preprint} \eprint{1006.0064})

\bibitem{2004CQGra..21R...1K}
{Karas} V, {Hur{\'e}} J~M and {Semer{\'a}k} O 2004 {\em Classical and Quantum
  Gravity\/} {\bf 21} R1--R51 (\textit{Preprint} \eprint{astro-ph/0401345})

\bibitem{2010MNRAS.404..545S}
{Semer{\'a}k} O and {Sukov{\'a}} P 2010 {\em \mnras\/} {\bf 404} 545--574
  (\textit{Preprint} \eprint{1211.4106})

\bibitem{2012MNRAS.425.2455S}
{Semer{\'a}k} O and {Sukov{\'a}} P 2012 {\em \mnras\/} {\bf 425} 2455--2476
  (\textit{Preprint} \eprint{1211.4107})

\bibitem{2013MNRAS.436..978S}
{Sukov{\'a}} P and {Semer{\'a}k} O 2013 {\em \mnras\/} {\bf 436} 978--996
  (\textit{Preprint} \eprint{1308.4306})

\bibitem{2015MNRAS.451.1770W}
{Witzany} V, {Semer{\'a}k} O and {Sukov{\'a}} P 2015 {\em \mnras\/} {\bf 451}
  1770--1794 (\textit{Preprint} \eprint{1503.09077})

\bibitem{2015llg..book..391M}
{Morris} M~R 2015 {\em {Manifestations of the Galactic Center Magnetic Field
  (Springer International Publishing Switzerland)}\/} p 391

\bibitem{2015Sci...350.1242J}
{Johnson} M~D, {Fish} V~L, {Doeleman} S~S, {Marrone} D~P, {Plambeck} R~L,
  {Wardle} J~F~C, {Akiyama} K, {Asada} K, {Beaudoin} C, {Blackburn} L,
  {Blundell} R, {Bower} G~C, {Brinkerink} C, {Broderick} A~E, {Cappallo} R,
  {Chael} A~A, {Crew} G~B, {Dexter} J, {Dexter} M, {Freund} R, {Friberg} P,
  {Gold} R, {Gurwell} M~A, {Ho} P~T~P, {Honma} M, {Inoue} M, {Kosowsky} M,
  {Krichbaum} T~P, {Lamb} J, {Loeb} A, {Lu} R~S, {MacMahon} D, {McKinney} J~C,
  {Moran} J~M, {Narayan} R, {Primiani} R~A, {Psaltis} D, {Rogers} A~E~E,
  {Rosenfeld} K, {SooHoo} J, {Tilanus} R~P~J, {Titus} M, {Vertatschitsch} L,
  {Weintroub} J, {Wright} M, {Young} K~H, {Zensus} J~A and {Ziurys} L~M 2015
  {\em Science\/} {\bf 350} 1242--1245 (\textit{Preprint} \eprint{1512.01220})

\bibitem{1998ApJ...508L..61L}
{Lo} K~Y, {Shen} Z~Q, {Zhao} J~H and {Ho} P~T~P 1998 {\em \apjl\/} {\bf 508}
  L61--L64 (\textit{Preprint} \eprint{astro-ph/9809222})

\bibitem{1998A&A...335L.106K}
{Krichbaum} T~P, {Graham} D~A, {Witzel} A, {Greve} A, {Wink} J~E, {Grewing} M,
  {Colomer} F, {de Vicente} P, {Gomez-Gonzalez} J, {Baudry} A and {Zensus} J~A
  1998 {\em \aap\/} {\bf 335} L106--L110

\bibitem{2008Natur.455...78D}
{Doeleman} S~S, {Weintroub} J, {Rogers} A~E~E, {Plambeck} R, {Freund} R,
  {Tilanus} R~P~J, {Friberg} P, {Ziurys} L~M, {Moran} J~M, {Corey} B, {Young}
  K~H, {Smythe} D~L, {Titus} M, {Marrone} D~P, {Cappallo} R~J, {Bock} D~C~J,
  {Bower} G~C, {Chamberlin} R, {Davis} G~R, {Krichbaum} T~P, {Lamb} J, {Maness}
  H, {Niell} A~E, {Roy} A, {Strittmatter} P, {Werthimer} D, {Whitney} A~R and
  {Woody} D 2008 {\em \nat\/} {\bf 455} 78--80 (\textit{Preprint}
  \eprint{0809.2442})

\bibitem{2000ApJ...528L..13F}
{Falcke} H, {Melia} F and {Agol} E 2000 {\em \apjl\/} {\bf 528} L13--L16
  (\textit{Preprint} \eprint{astro-ph/9912263})

\bibitem{2009A&A...496...77F}
{Falcke} H, {Markoff} S and {Bower} G~C 2009 {\em \aap\/} {\bf 496} 77--83
  (\textit{Preprint} \eprint{0901.3723})

\bibitem{2012MNRAS.421.1517D}
{Dexter} J, {McKinney} J~C and {Agol} E 2012 {\em \mnras\/} {\bf 421}
  1517--1528 (\textit{Preprint} \eprint{1109.6011})

\bibitem{2010MNRAS.403L..74K}
{Kato} Y, {Miyoshi} M, {Takahashi} R, {Negoro} H and {Matsumoto} R 2010 {\em
  \mnras\/} {\bf 403} L74--L78 (\textit{Preprint} \eprint{0906.5423})

\bibitem{2010A&A...510A...3Z}
{Zamaninasab} M, {Eckart} A, {Witzel} G, {Dovciak} M, {Karas} V, {Sch{\"o}del}
  R, {Gie{\ss}{\"u}bel} R, {Bremer} M, {Garc{\'{\i}}a-Mar{\'{\i}}n} M,
  {Kunneriath} D, {Mu{\v z}i{\'c}} K, {Nishiyama} S, {Sabha} N, {Straubmeier} C
  and {Zensus} A 2010 {\em \aap\/} {\bf 510} A3 (\textit{Preprint}
  \eprint{0911.4659})

\bibitem{2010PhRvD..82h4034F}
{Frolov} V~P and {Shoom} A~A 2010 {\em \prd\/} {\bf 82} 084034
  (\textit{Preprint} \eprint{1008.2985})

\bibitem{2014arXiv1410.1663P}
{Piotrovich} M~Y, {Gnedin} Y~N, {Silant'ev} N~A, {Natsvlishvili} T~M and
  {Buliga} S~D 2014 {\em arXiv e-prints\/} arXiv:1410.1663 (\textit{Preprint}
  \eprint{1410.1663})

\bibitem{Bar-Car-Haw:1973:CMP:}
{Bardeen} J~M, {Carter} B and {Hawking} S~W 1973 {\em Communications in
  Mathematical Physics\/} {\bf 31} 161--170

\bibitem{1977MNRAS.179..433B}
{Blandford} R~D and {Znajek} R~L 1977 {\em \mnras\/} {\bf 179} 433--456

\bibitem{1985ApJ...290...12W}
{Wagh} S~M, {Dhurandhar} S~V and {Dadhich} N 1985 {\em \apj\/} {\bf 290} 12--14

\bibitem{2018MNRAS.478L..89D}
{Dadhich} N, {Tursunov} A, {Ahmedov} B and {Stuchl{\'{\i}}k} Z 2018 {\em
  \mnras\/} {\bf 478} L89--L94 (\textit{Preprint} \eprint{1804.09679})

\bibitem{1975ARA&A..13..381E}
{Eardley} D~M and {Press} W~H 1975 {\em \araa\/} {\bf 13} 381--422

\bibitem{PhysRevD.98.123002}
Levin J, D'Orazio D~J and Garcia-Saenz S 2018 {\em Phys. Rev. D\/} {\bf 98}(12)
  123002 \urlprefix\url{https://link.aps.org/doi/10.1103/PhysRevD.98.123002}

\bibitem{2017CoSka..47..124K}
{Karas} V, {Kop{\'a}{\v c}ek} O, {Kunneriath} D, {Zaja{\v c}ek} M, {Araudo} A,
  {Eckart} A and {Kov{\'a}{\v r}} J 2017 {\em Contributions of the Astronomical
  Observatory Skalnate Pleso\/} {\bf 47} 124--132 (\textit{Preprint}
  \eprint{1705.09820})

\bibitem{2018PhRvD..98h4055K}
{Kop{\'a}{\v{c}}ek} O, {Tahamtan} T and {Karas} V 2018 {\em \prd\/} {\bf 98}
  084055 (\textit{Preprint} \eprint{1810.04220})

\end{thebibliography}


\end{document}